# Reversible Transition between Superoleophobic and Superoleophilic States on Titania Coated Substrates by UV Irradiation

*JITESH BARMAN,[a] SUMIT KUMAR MAJUMDER[a] AND KRISHNACHARYA KHARE, [a]\**

*Email: kcharya@iitk.ac.in

[a]DEPARTMENT OF PHYSICS, INDIAN INSTITUTE OF TECHNOLOGY KANPUR, KANPUR - 208016, INDIA

ABSTRACT

We demonstrate that tunable superoleophobic surfaces fabricated by a simple spin and spray coating methods of titania on silicon (Si) wafers and stainless steel (SS) mesh, which possess a hierarchical re-entrant structure consisting of nano meter sized particles on top of micron sized particle, are able to induce superoleophobicity on an oleophilic self assembled monolayer of 1H,1H,2H,2H-perflurodecyltrichlorosilane (FDTS). Though comparison between different coating methods, we show that spray coating on Si substrate and SS mesh can repel lower surface tension liquids than spin coating on Si substrates due to formation of half spherical micro-particles on the spray coated substrates which is confirmed by FESEM images. Subsequently, superoleophobic surfaces changes its super repellent property to complete wetting of any liquids under UV illumination by decomposing the FDTS molecules via photo catalytic property of titania particles. The superoleophobic property was regained by annealing followed by grafting of FDTS on the UV treated surface and the total process is repeatable over number of cycles.

1. INTRODUCTION

The discovery of natural surfaces (lotus leaf effect) with super repellent property for water i.e. water contact angle close to 180° lead to extensive theoretical as well as experimental research on artificial superhydrophobic surfaces with high roughness. Over the decade many research groups have demonstrated a variety of potential applications of superhydrophobic surfaces from our daily life to industry like self cleaning[1-5], antifreezing[6], anticorrosion[7-9], antisticking of snow[10] and anti biofouling[11]. Inspired from superhydrophobic surfaces exist in nature like plant leaves[12], legs of water strider[13], wings of butterfly[14], feathers of birds[15] etc. people have fabricated superhydrophobic surfaces in laboratory. The understanding of the two key parameters (surface energy and surface roughness) is necessary to fabricate superhydrophobic surfaces. The apparent contact angle of a liquid on a rough surface according to the Wenzel model[16] and Cassie model[17] respectively are,

$$\cos \theta^* = r \cos \theta \tag{1}$$

$$\cos \theta^* = f \cos \theta + f - 1 \tag{2}$$

Where θ is the intrinsic contact angle or the Young's contact angle, $r$ is the roughness factor and $f$ is the fraction of the wet surface. Cassie-Baxter model predicts the super repellence for water with a very low roll of angle of surface with hierarchy in roughness (micro and nano)[18, 19] and intrinsic contact angle of water > 90° . However, superoleophobic surfaces which is super repellent for oils or liquids with low surface tension (< 35 mN/m), are very difficult to fabricate due to the fact that these liquids show contact angle <90° with almost all known materials.[20, 21] Hence the surface roughness and the fraction of the wet surface area both should be modified in such a way that can hold a low surface tension liquid with low contact angle to prepare a superoleophobic surface. Tuteja *et al.*[22] was the first group to realize the requirement of 're-entrant' curvature or 'overhang' property in the surface roughness for preparation of superoleophobic surface with oil contact angle in the Cassie regime. Improvising these surface properties people has fabricated superoleophobic surfaces in lab extensively.[22-32]

The reversible conversion between superhydrophobic states to superhydrophilic states can be achieved either by tuning surface energy by various types of external stimuli including optical radiation[33-35], heat[36], electric field[37, 38], surfactant[39] etc. or by changing surface roughness. Among these two, tuning the surface energy by external stimuli is way easier and can be controlled very precisely than tuning the surface roughness. However, switching between extreme wetting states for oils is very rare due to the low surface tension of the oils. The photo catalytic and the photo stimulated properties of the titania allows the conversion of the surface property from super-hydrophobic to super-hydrophilic and vice versa by UV irradiation and heating respectively. Also a FDTS coated surface shows unidirectional conversion of the surface property from hydrophobic to hydrophilic upon UV irradiation due to gradual decomposition of FDTS[40]. Although oleophobic surfaces, tunable super-hydrophobic surfaces have been fabricated but simultaneously tunable super-hydrophobic and oleophobic surface has not been observed due to the difference in polarity and surface tension of water and oil.

In this work, a time efficient and very simple process for fabrication of a superoleophobic surface by spray coating of titania on any smooth solid surface is demonstrated. The surface property can be tuned from superoleophobic to superoleophilic state upon UV illumination on substrates. Moreover the superoleophobicity can be recovered from superoleophilic state by heating and FDTS monolayer grafting on the surface.

## 2 EXPERIMENTAL SECTIONS

### 1. Surface modification of titania coatings:

Single side polished p-type doped Silicon (Si) wafer (University of wafer, USA) and stainless steel (SS) mesh (70 μm pitch distance with 30 μm wire diameter) cut into square pieces (2 cm X 2 cm) were used as substrates. The substrates were thoroughly cleaned in ethanol, acetone and toluene by ultrasonicating followed by plasma cleaning (Harrick plasma, USA). 6 wt% of titania nano particles (mean diameter ~21nm, Sigma Aldrich) were dispersed in isopropanol (Merk, India ) and put in ultrasonication for 20 minutes to prepare a uniform dispersion. The titania dispersion was then coated on the pre cleaned substrates by spin coating and spray coating methods. The spin coating parameters: 1000 RPM, acceleration time-30 sec and rotation time-60 sec leads to a uniform of coating of titania on smooth Si substrates. The pre cleaned Si and SS mesh were spray coated with parameters: pressure 2 kg/cm$^2$, distance 10 cm and duration-2 minutes. The titania coated substrates were subsequently annealed at 80°C for 20 minutes to evaporate the solvent from samples. The titania coating was rendered superoleophobic by grafting a monolayer of 1H,1H,2H,2H-perfluorodecyltrichlorosilane (FDTS) (Alfa Aesar, India) following the evaporation method. Later the samples were exposed to UV irradiation (wavelength ~254 nm and intensity~350 μW/cm$^2$) for 6 hrs to modulate the wettability of the coatings. To recover superoleophobicity the UV exposed samples were annealed at 180°C for 30 minutes and another monolayer of FDTS was grafted.

## 2. Characterization:

Field emission- scanning electron microscopy (FE-SEM), X-ray photoelectron spectroscopy (XPS) and Goniometer were used to characterize different surface properties of titania coated superoleophobic and superoleophilic substrates. The morphology of the titania coated samples were investigated by FE-SEM (Supra 400VP, Zeiss, Germany) at an accelerating voltage 10kV. The samples were not coated with an additional layer of metal during the FESEM imaging. The atomic percentage of different elements for different conditions was measured by X-ray photoelectron spectroscopy (PHI 5000 Versa Probe II, FEI Inc.).

Static contact angle of different liquids was measured by an optical contact angle goniometer (OCA- 35, Dataphysics, Gemany) using a fixed volume (5μl) of sessile drop. For each sample, a minimum of five data points were taken at different spots on the sample.

## 4. RESULTS AND DISCUSSION

Static contact angle (CA) of DI water, various oils (olive oil, hexa decane, n-decane) and ethanol solution were measured on the as prepared spin coated samples using the goniometer. The CA of DI water (surface tension ($\gamma$= 72 dynes/cm ), olive oil($\gamma$=32 dynes/cm), hexadecane ($\gamma$= 27.5 dynes/cm ) and n-decane (($\gamma$= 23.4 dynes/cm) at room temparature were around 160° 135°, 60°, and 45° respectively. The measurement implies that the spin coated surface is purely super-hydrophobic and oleophobic (for oils upto $\gamma > 30$ dynes/cm) in nature. Doing measuments with oils on this surfaces is difficult as the oils get stick to substrates results in altering the chemical composition. For CA measurements on spin coated substrates 0- 60 wt% ethanol solutons were used as it is easy to prepare a solution of required surface tension in the range of 72-20 dynes/cm. Furthermore, these solutions evaporate quickly without leaving a trace on the surface unlikely oils and do not alter the chemical composition of the surface. The CA of various wt% aquaous solution of etahnol as a function of surface tension is shown in figure 1.

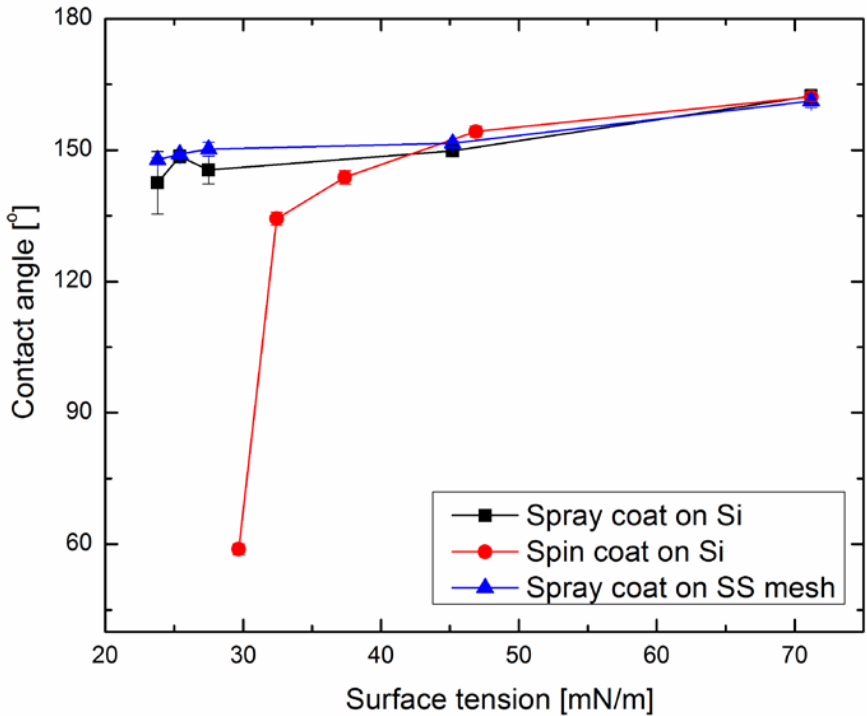

*Figure 1. Contact angle as a function of surface tension of liquidsfor different coating methods. Red circles for spin coated Si substrate, black squares are for spray coated Si substrate, and blue triangles are for spray coated SS mesh.*

The spin coated substrates show CA around 160° for DI water which confirms that the surface is super hydrophobic. As the surface tension of the test liquid is decreased, the contact angle decreases to 130°- 135° for ranges γ=30-35 dynes/cm confirms the oleophobicity of spin coated substrates and further decrease in surface tension shows a sharp decrease in contact angle about 50°-60°.

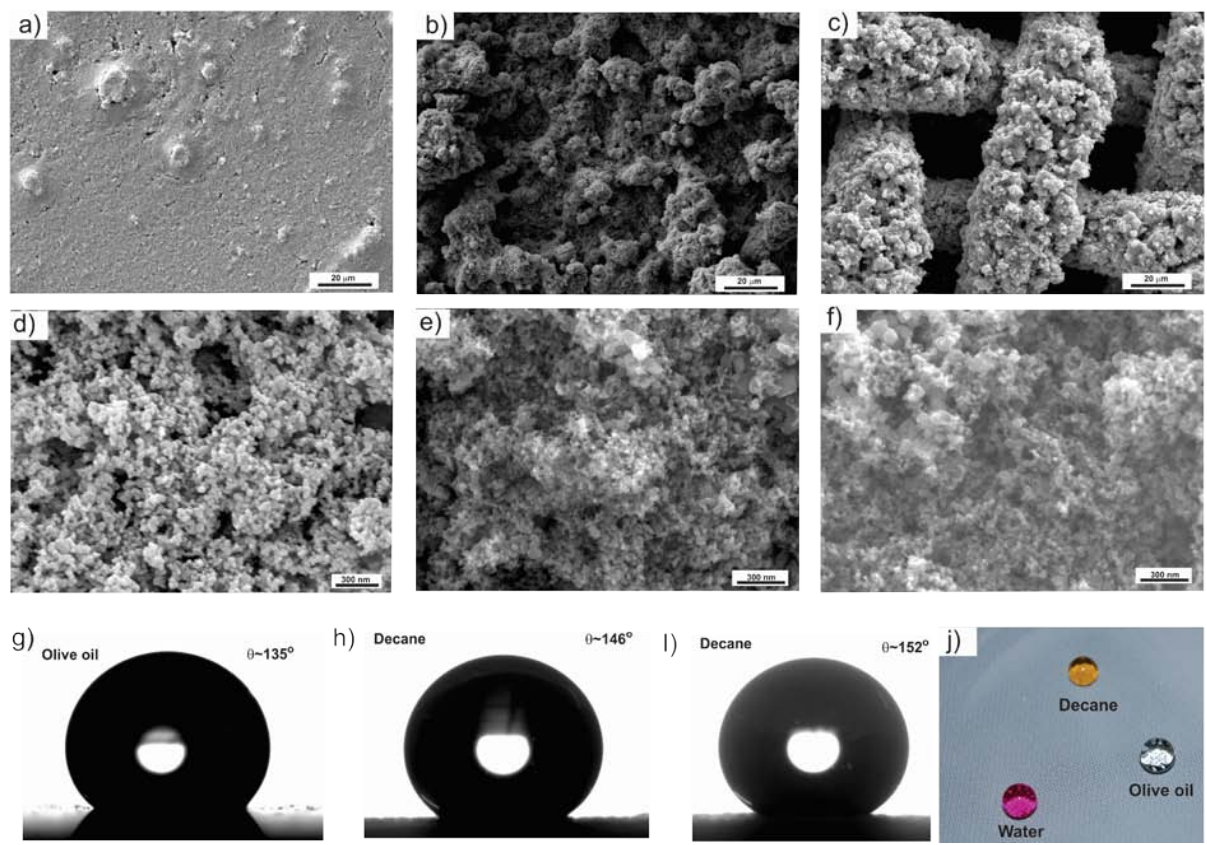

*Figure2. Superoleophobicity is realized on titania coated Si and SS mesh by different coating method. FESEM images with 20 mm scale bar of titania surface coated by (a)spin coating method on Si substrate, (b)spray coating method on Si substrate and (c)spray coating method on SS mesh.(d),(e),and (f)are the high magnification FESEM images with scale bar 300 nmof spin coated Si substrate, spray coated Si substrate, spray coated SS mesh respectively which show the presence of reenetrancity in nano particles.(g) A olive oil drop on a spin coated Si substrates shows contact angle about 135°. (h) A decane drop with contact angle 146° on spray coated Si subatrate. (i) A Decane drop with colored dye does not penetrate through the sray coated SS mesh.(j)Spherical drops of wtaer(with colored dye), decane(with colored dye)and olive oil on a superoleophobic SS mesh.*

The spray coated Si susbtrates and SS meshes both similarly show superhydrophobicity as well as super repellenncy towards very low surface tension liquids. The spray coating of titania pushes the limit of superoleopbocity (CA close to 150°) to the liquids with surface tension upto 23 mN/m as shown in figure 1. Figure 2 (h) and (i) show a decane drop on top of spray coated Si substrate and SS mesh with contact angle 146° and 150° respectively. The stainless steel (SS) mesh shows slightly better superoleophibicity than Si substrates which can be seen from figure 1. The spin coated substrates only show oleophobicity upto certain range of surface tension which is higher than the spray coated substrates which confirms the satement that the spray coating enhance the superoleophobicity substrates.

Surface morhology and the surface energy of the coatings are very important factors for fabrication of superolophobic surfaces. Nano and micro scale roughness with reentrant curvature are necessary in surface morphology as well as presence of a low surface encergy layer which enhances the contact angle than the bare rough surface. Figure 2 (a-f) shows the SEM image of the morphology of $TiO_2$ coated Si and SS mesh substrates by spin and spray coating. An important oberservation from figure 2(a) is that during spin coating of the $TiO_2$ dispersion the nano particles form a microscpoically less rough surface with no reentrant geometry on it . High magnified image of spin coated substrates in figure2 (d) shows that nano particles with particle size around 20 nm makes a porous cluster. The spherical shape of the nano particles gives the necessary re-entrant curvature in nano meter sized surface roughness. This hierachical roughness with FDTS coating leads to surperhydrophocity. The reentrant curveture present in the nano roughness results in oleophobicity of the spin coated substrates. These substrates do not show oleophobicity with very low surface tension (< 30 mN/m) due to the lack of reentrant property in the micron sized roughness.[41] Figure 2 (b) and 2(e) show FE-SEM image of the morphology of spray coated Si substrate. Spray coating of $TiO_2$ dispersion on Si substrate results in formation of micron sized bead type structure with spherical shape. The magnified image of the surface shows the porous cluster of nano particles with spherical shape which is similar to the spin coated Si substrate. Spray coated substrates show superoleophobicity for very low surface tension liquids due to presence of reentrant property in highly dense the micro as well as nano structure. From figure 2 (c) one can clearly see that the pray coating of titania on SS mesh covers all the cylindrical wire of the mesh leaving the gap between the two wires. Another important observation of morphology shows that the titania particles make clusters and gives rise to flower like structure on the wires of mesh. The mesh wires adds more micron sized roughness with flower type structure. The hierarchy in the surface roughness and the presence of renenrtant property in the surface morphology leads to better super repellent to liquids with very low surface tension than the spray coated Si substrates.

There are mainly two methods to tune the surface property from super oleophobic to superoleophilic; one is to change the surface roughness and other is to change the surface energy of the surface by some external stimuli. In our system the later one is way more convenient and controlled than the other due to the photocatalytic property of the titania nano particles. Cui et al. have shown that upon UV irradation the FDTS gradually decomposes due to photocatalyic action of $TiO_2$ surface[40]. The spray coated Si substrates were kept under UV irradiation (256 mW power) for 6 hrs. for the complete or partial change of surface property from superoleophobic to superoleophilic. Static CA of DI water as well as different oil were measured on this surface after UV exposure. Static CA for DI water on the UV exposed substrates show complete wetting as shown in figure 3(a). An olive oil drop similarly shows complete wetting on the exposed substrates confirming that the substrate has changed its state from super repellent to super wetting irrespective of there polarity. Then the UV exposed substrates were annealed at 185° for 30 minutes . Thereafter, the substrates were placed in vaccum for single layer molecule of FDTS grafting which restore the superoleophobicity of the substrate. The static CA of water and olive oil are around 160 and 150 respectively after the surface modification as shown in figure 3 (a). This reversible tuning between super repellent and wetting states is captured in figure 3(b) for olive oil over a number of cyles.

FESEM images (not shown in this article) of the UV irradiated substrate confirm that the surface morphology of titania particles did not change upon UV exposure. The surfaces property undergoes transition from superoleophobic to superoleophilic due to the decomposition of FDTS mono layer under UV irradiation. Decomposition of FDTS monolayer is not a direct consequence of UV irradiation but rather an idirect consequence of photocatalytic effect of titania nano particles. To confirm the above mentioned fact, a smooth Si wafer was grafted with FDTS monlayer and the CA of water and other oils was measured before and after the UV irradiation. The measured CA remained same in both the cases which is the result of non degradation of FDTS monlayer under UV irradiation. Titania is photocatalytic semiconductor with band gap 3.26 eV and titenium is in +4 ($Ti^{4+}$ state) state. Due to absorption of UV irradiance $Ti^{4+}$ undergoes a transition to $Ti^{3+}$ state and it comes back to its normal 4+ state after annealing at 180°C.

$$Ti^{4+} \xrightarrow{UV\ irradiation} Ti^{3+}$$

$$Ti^{3+} \xrightarrow[Heating]{} Ti^{4+}$$

The transition from $Ti^{4+}$ state to $Ti^{3+}$ decomposes the FDTS molecules.The heating of the samples after UV treatment allows the convertion of the electronic state of $Ti^{3+}$ to $Ti^{4+}$. FDTS layer deposition helps to change the surface energy of the surface thus the surface regains its superolephobicity.

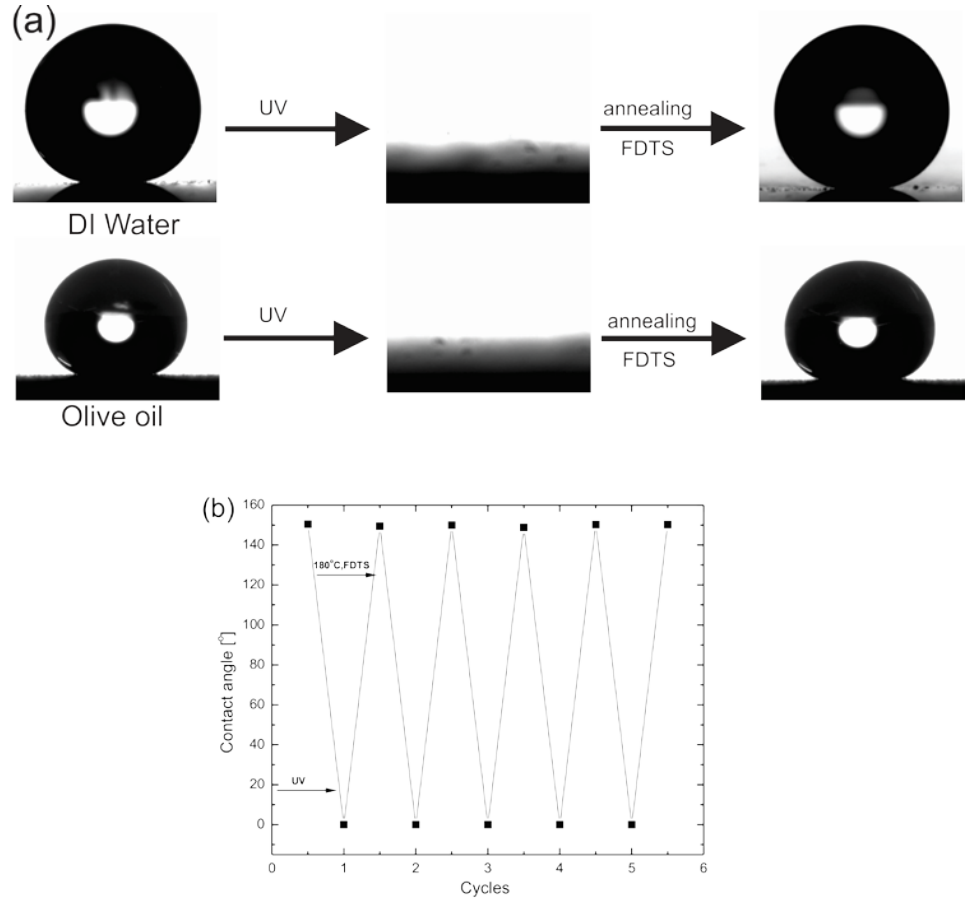

Figure 3.Wettability transition from superoleophobic to superoleophilic on spray coated Si subsrates under UV irradiance. (a) Water and Olive oil drop on a superoleophobic surface under goes a transtion to spreading state after UV irradaition. Surface recovers its oleophobicity due to annealing followed by FDTS grafting. (b) Static CA of olive oil on spray coated Si substrate for different cycles.

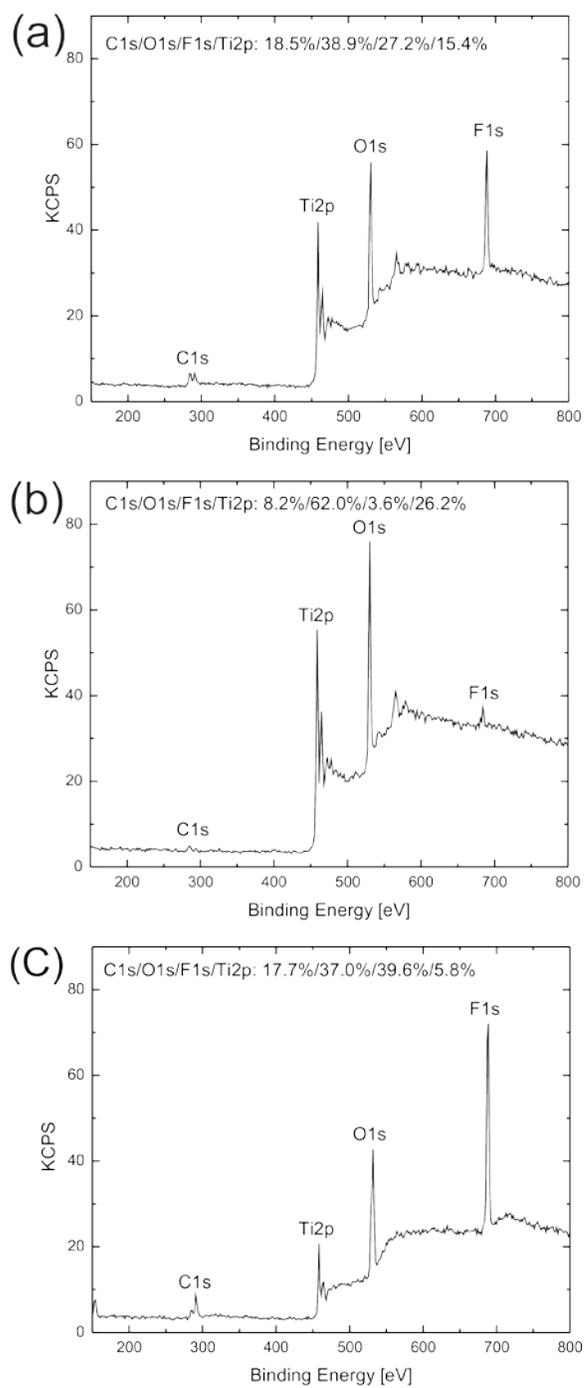

*Figure 4: XPS spectra of different elements of (a) UV untreated sample, (b)UV treated sample and (c) UV treated, annealed & FDTS grafted sample. The inset shows the percentage of different elements.*

To further verify the reversible transition between superoleophobic and superoleophilic states, XPS analysis of the substrates with UV untreated, UV treated and UV treated followed by annealed & FDTS grafted were investigated. The variation of atomic concentration of different elements corresponding to different bonds were compared in figure 4 and table 2. Figure 4 (a) shows XPS spectra of UV untreated substrate where four peaks were derived from F 1s (corresponding to C-F bonds of FDTS; 688.8 eV), O 1s (corresponding to O-Ti bonds of titania; 530.4 eV), Ti 2p (corresponding to O-Ti bonds of titania; 458.4 eV), and C 1s (corresponding to C-F bonds of FDTS; 290.4 eV). The atomic percentage analysis shows that the surface has 27.2 % of flourine corresponds to FDTS, 38.9 % of oxygen corresponds to $TiO_2$. This idicates that surface has a low surface energy coating of FDTS which leads to superoleophobicity.

Table 2: Comparison of atomic percentage of different elements of samples at different conditions

|  | F % | O% | C % |
|---|---|---|---|
| As prepared sample | 27.2 | 40 | 18 |
| Uv treated sample | 2.7 | 62 | 8.2 |
| Annealed grafted | 40 | 37 | 17 |

The UV exposure for 6hrs on surface shows a decrease in atomic percentage of flourine (3.6%) and carbon (8.2%) as shown in table 2 which is consistent with XPS spectra (figure 4b) where a reasonable decrease in the peak intensity of flourine (688.8 eV) and carbon (290.4 eV) can be seen. Also one can clearly see that there is an increase in atomic percentage of oxygen (62%) corresponding Ti-O bonding which is a direct consequence of degradation of FDTS monlayer and traping of electron from environment in defect state of titania. As a consequence of decomposition of oleophobic FDTS and increase in oleophilic component titania in the surface leads to zero CA (spreading) of oils on the surfaces. Whereas after annealing at 180°C and regrafting of FDTS on the UV treated samples show similar characteristic like the UV untreated substrates. The atomic percentage (39.6%) as well as the intesnsity of peak (530.4 eV) of oxygen decreases compared to the UV treated samples. From table 2 and figure 4c, it can be seen that the atomic percentage and intensity of flourine peak (688.8 eV) increases drastically compared to UV untreated samples. The increase in low surface energy FDTS and decrease in oleophilic component titania on surfaces leads to recovery of superoleophobic state of the surface.

5. CONCLUSION

In summary, we have demonstrated a very simple method of fabrication of superoleophobic surfaces by spray and spin coating of titania nano particles followed by FDTS grating. Out of these two coating methods, spray coating shows better superoleophobicity. The presence of re-entrant curvature property in nano roughness and micro roughness as well as presence of the FDTS single layer is the reason for the superoleophobicity of the surface. Thereafter, the surface properties can be tuned from superoleophobic to superoleophilic by UV irradiation. Furthermore, the superoleophobic property can be regained through annealing and FDTS grafting of the UV treated substrate. The total process is reversible and repeatable for 20 times.